
\raggedbottom


\def\nex{\par\noindent\hang}

\null
\centerline{\bf UNIFICATION OF THE NEARBY AND PHOTOMETRIC}
\vskip 2mm
\centerline{\bf STELLAR LUMINOSITY FUNCTIONS}
\vskip 10mm
\par
\centerline{\bf Pavel Kroupa}
\vskip 10mm
\centerline{Astronomisches Rechen-Institut}
\vskip 2mm
\centerline{M{\"o}nchhofstra{\ss}e~12-14, D-69120~Heidelberg, Germany}
\vskip 15mm
\centerline{e-mail: S48 @ IX.URZ.UNI-HEIDELBERG.DE}


\vskip 140mm

\centerline{ApJ (part 1), in press}

\vfill\eject

\centerline{\bf Abstract}
\vskip 5mm
\noindent
We introduce a model Galactic field low-mass stellar
population that has a proportion of binary systems as observed, with a
mass ratio distribution consistent with observational constraints. The
model single
star and system luminosity function agrees with the nearby and the Malmquist
corrected photometric luminosity function, respectively. We tabulate the model
luminosity functions
in the photometric V-, I- and K-bands, and in bolometric magnitudes.
Unresolved binary systems
are thus a natural explanation for the
difference between the nearby and photometric luminosity functions. A local
overdensity of faint stars needs not be postulated to account for the
difference, and is very unlikely.
We stress that the nearby luminosity
function can only be used to weakly constrain the stellar mass function below
$0.5\,M_\odot$, because of the small sample size. The photometric luminosity
function can only be used to put lower limits on the stellar mass
function because most binary systems are not resolved. However, taken together
the nearby and photometric
stellar luminosity function data do not imply a mass function with a peak at a
mass of $0.2-0.3\,M_\odot$. Instead, the data are consistent with a power-law
mass function between 0.08$\,M_\odot$ and $0.5\,M_\odot$. We urge researchers
to only use star
count data that are properly corrected for {\it all} contributions to cosmic
scatter, photometric uncertainties, and
unresolved binaries, and to be aware of the severe limitations of theoretical
mass--luminosity relations for low mass stars, when drawing conclusions about
structure in the stellar mass function.

\vskip 5mm
\noindent {\it Subject headings:} stars: low-mass -- stars: luminosity
function, mass function, mass--luminosity relation

\vfill\eject

\vskip 0.1in
\vskip 24pt
\noindent{\bf 1 INTRODUCTION}
\vskip 12pt
\noindent
The astrophysical importance of the stellar mass function for stars less
massive than one solar mass
has been stressed many times and much effort has gone into constraining its
shape. Interest is inspired from the cosmological and Galactic dynamics fields
(how much baryonic mass is stored in faint low mass stars?) as well as
from the star formation field (does the initial mass function vary among star
formation regions and is there a lower mass limit below which no `stars'
form?). The stellar mass function cannot be observed directly but can be
estimated from the stellar luminosity function. Unfortunately, observational
attempts at constraining the low mass stellar luminosity function have led to
rather discordant results.

The `nearby' luminosity function, $\Psi_{\rm near}$, constructed from stars
within 5--20~pc distance
of the Sun using trigonometric parallax measurements to estimate stellar number
densities has a {\it significantly} larger stellar number density
at $M_{\rm v}>13$ than stellar luminosity functions estimated from
low spatial resolution but deep (100--200~pc distance) magnitude limited
surveys (Kroupa 1995a). We refer to these as
`photometric' luminosity functions, $\Psi_{\rm phot}$, because distances are
estimated using
photometric parallax. Photometric luminosity functions need to be carefully
corrected for systematic bias which results because stars of a given colour do
not all have the same absolute magnitude owing to different metal abundances,
unresolved binary components and different ages. The resulting dispersion of
absolute magnitude at a particular colour in the colour--magnitude
diagram is referred to as cosmic scatter.
The imposed magnitude limit leads to an
apparently larger stellar number density and an apparently brighter stellar
sample. This effect is referred to as Malmquist bias.
In Kroupa (1995a) we discuss this issue in greater detail and we combine four
independent
photometric luminosity functions to estimate the parent Malmquist corrected
photometric
luminosity function which is based on a sample of 448~stars. Applying a number
of different statistical tests we establish that $\Psi_{\rm near}$ and
our estimate of the true Malmquist
corrected photometric luminosity function, ${\overline\Psi}_{\rm phot}$, are
significantly different at $M_{\rm V}>13$. Both luminosity functions are
tabulated in table~2 of Kroupa (1995a).

The difference has been claimed by Kroupa, Tout \& Gilmore (1991, 1993) to
result from unresolved binary systems in the low-spatial resolution
photographic surveys used to estimate $\Psi_{\rm phot}$. They find that the
nearby and photometric luminosity functions can be understood to stem from the
same underlying stellar mass function provided it is approximated by two
power-law segments between 0.08~and $1.1\,M_\odot$. However, Reid (1991)
concludes that unresolved binary systems cannot account for the difference.
Furthermore, Tinney (1993) uses the estimate of the photometric luminosity
function from
the recent {\it precise} large-scale photographic survey of Tinney, Reid \&
Mould (1993) to address this issue. Tinney (1993) and Reid (1994) conclude (i)
that the nearby
and Tinney's photometric luminosity function do not differ and (ii) that the
stellar mass
function has a significant maximum at a mass of about $0.2-0.3\,M_\odot$, thus
contradicting the results of Kroupa et al. (1991, 1993).

In Kroupa (1995a) we resolve point~(i) by noting that the conclusion by
Tinney (1993) and Reid (1994) rests upon the comparison of
the nearby luminosity function with the
photometric luminosity function in bolometric magnitudes which has not been
corrected for Malmquist bias. Malmquist bias in
Tinney's photometric luminosity function leads to significant overestimation of
stellar number densities. Comparing Tinney's photometric luminosity function
corrected to first order for Malmquist bias by Tinney et al. (1993)
with $\overline{\Psi}_{\rm phot}$ we find good agreement between the two.
However, point~(ii) remains unresolved, and we take the
suggestion by Tinney (1993) and Reid (1994) seriously, because the survey
of Tinney et al. (1993) significantly improves our estimate of the photometric
luminosity function being based on about 3500~stars.

In this paper we study the possible reasons for the difference between the
nearby and Malmquist corrected photometric luminosity functions. We also
critically investigate the evidence for structure in the Galactic field stellar
mass function claimed by Tinney (1993) and Reid (1994).

In Section~2 we contemplate whether a local stellar overdensity may be
responsible for the significant difference between $\Psi_{\rm near}$ and
$\overline{\Psi}_{\rm phot}$. In Section~3 we critically examine
the models of Reid (1991) and Kroupa et al. (1993), and in Section~4 we
introduce
a realistic model of the dynamical properties of stellar systems in the
Galactic disc and compare the model with the observed luminosity functions. In
Section~5 we discuss the stellar mass function, and Section~6 contains our
conclusions.

\vfill\eject

\nobreak\vskip 10pt\nobreak
\noindent{\bf 2 A LOCAL STELLAR OVERDENSITY?}
\nobreak\vskip 10pt\nobreak
\noindent
Perhaps near the Sun there are {\it significantly} more stars with $M_{\rm
V}>13$ than at a distance of $100-200$~pc (see section~4.3 in Kroupa 1995a).
Photometric luminosity
functions in directions towards the north and south Galactic poles, and in
directions not perpendicular to the Galactic plane, show that the
shape of the distribution
of stars with luminosity does not vary significantly with direction for
$M_{\rm V}<16.5$ (fig.~1 in Kroupa 1995a; fig.~5 in Stobie et al. 1989; fig.~20
in Tinney
et al. 1993). The different normalisation can readily be accounted for by
correcting for a density gradient perpendicular to the Galactic midplane.
This hypothesis would imply
that the sun is situated in a bubble containing significantly more faint stars
than the Galactic field.

The velocity dispersion of the old Galactic disc
population is about 50~km~sec$^{-1}$, and within 4~Myrs such a population would
have dispersed over a distance of 200~pc. The nearby stars with $13<M_{\rm
V}<16.5$ that define $\Psi_{\rm near}$ have kinematics typical of an old
population
(section~10.3 in Kroupa et al. 1993). A {\it significant} local population
of young faint stars therefore cannot account for the
difference between the nearby and photometric luminosity functions, as also
stressed by Reid (1991) in his section~4. Thus, in
order to explain the difference between the nearby and photometric luminosity
functions with the ``overdensity hypothesis'' we need to {\it postulate} a very
special (i.e. only faint old stars) {\it and} an extremely short lived
constellation of stars.

Clearly this is not a satisfying solution to the problem.

\nobreak\vskip 10pt\nobreak
\noindent{\bf 3 BINARY STARS}
\nobreak\vskip 10pt\nobreak
\noindent
If in the galactic disk 50-60~per~cent of all ``stars'' are binary systems and
if the mass ratio distribution is biased towards
non-equal masses, then straightforward logics implies that the
system
luminosity function decreases significantly with decreasing luminosity when
compared to the single star luminosity function. This has been shown by Kroupa
et al. (1991), who have also demonstrated that one
mass function can `unify' both the nearby and photometric luminosity
functions. This approach increases the statistical weight of the low-mass data.

Observational evidence indicates that the above proportion of binary systems is
about correct, and that there are more main sequence binaries with low-mass
secondaries than with high mass secondaries.
Photographic surveys have a resolution of 3--6~arc~sec (Reid 1987) which, at a
distance of 100~pc, corresponds to 300--600~AU. For a binary system with total
mass of $1.3\,M_\odot$ this corresponds to log$_{10}P=$6.2--6.7, where $P$ is
the
orbital period of the binary system in days. This range of orbital periods lies
well beyond the maximum at log$_{10}P\approx5$ in the period distribution of
late-type binaries (see Kroupa 1995b and references therein). Thus virtually no
binary systems can be resolved by photographic techniques.

\nobreak\vskip 10pt\nobreak
\noindent {\bf 3.1 They don't matter?}
\nobreak\vskip 10pt\nobreak
\noindent
Reid (1991) adopts a similar approach to that of Kroupa et al. (1991) but
instead of drawing stars from a mass function he
draws single stars from an adopted model of the nearby luminosity
function. He redistributes stars in the observed nearby luminosity function for
`aesthetic reasons' to give his preferred model which is flat for $M_{\rm
V}>12$ and on which his models A--J and M,N are based. He also considers a
model single
star luminosity function which has a pronounced maximum at $M_{\rm V}\approx12$
(his models~K and~L taken from Kroupa et al. 1991).

Some of the stars Reid (1991) combines to binary
systems and changes the photometric
properties of the resulting population to account for Gaussian cosmic
scatter (his adopted model single star luminosity function is thus an ideal
model without measurement errors of a single metallicity and single age stellar
population). The resulting model photometric luminosity function he compares
with his fit, $\Psi_{\rm Reid,phot}$, to
the logarithmic data presented in fig.~5 in Stobie et al. (1989) which are not
corrected for Malmquist bias.
We note that his adopted observed photometric luminosity function appears in
his fig.~9
stretched for unspecified reasons (thus enhancing the decay at $M_{\rm V}>12$).
For example, in his table~1
we find $\Psi_{\rm Reid,phot}=2.0\times10^{-3}$~stars pc$^{-3}$ mag$^{-1}$ at
$M_{\rm V}=16$. In fig.~9 `D' this datum appears as $\Psi_{\rm
phot}=1500$~stars mag$^{-1}$ which implies a scaling factor of 750. Thus we
would expect $\Psi_{\rm Reid,phot}=12.50\times10^{-3}$~stars pc$^{-3}$
mag$^{-1}$ at $M_{\rm V}=12$ (his table~1) to appear as $\Psi_{\rm
phot}=9375$~stars mag$^{-1}$ in his fig.~9 `D'. Instead we find $\Psi_{\rm
phot}=11300$ stars mag$^{-1}$. It is important to bear this in mind because
his conclusions are based on an eye-ball comparison of the shapes of the
luminosity functions.

Reid adopts a number of models for the binary star population based on stellar
data within a distance of 5--10~pc. He assumes his binary population consists
of `wide' binaries (components chosen at random from his adopted model
single-star
luminosity function) and `equal-mass' binaries (components restricted to lie
within 2~mag of the luminosity of the primary -- note that for stars with
$5\le M_{\rm V}\le12$ this corresponds
to a mass range of $0.17\,M_\odot$ using his adopted linear mass--$M_{\rm V}$
relation and to a mass range of about $0.1\,M_\odot$ or less for fainter
stars).

His models assume a uniform spatial distribution of stars and a
Gaussian cosmic scatter independent of absolute magnitude.

{}From his fig.~9 Reid (1991)
concludes that his model photometric luminosity functions make a poor eye-ball
comparison
with his adopted observational photometric luminosity function. He rejects the
hypothesis that unresolved binary systems account for the difference between
the nearby and photometric luminosity functions, putting greatest weight on his
`favorite' models F, G and H (30--50~per cent binary population, of which
50~per cent are wide binaries and 50~per cent have equal-mass components).

His conclusion does not rest
on a statistical analysis, and so its confidence cannot be assessed.
Below we show that observational constraints contradict these models.

It is useful at this point to consider why Stobie et al. (1989)
also conclude that binary systems cannot explain the difference between the
nearby and photometric luminosity functions. They use the sample of 60~stars
within 5.2~pc of the sun. This sample consists of 45~systems of which
32~are single
stars. From this sample they obtain the single star and system luminosity
function and find no eye-ball difference between the two. In this case, only
5--7~stars per magnitude bin are in the single star luminosity function. Thus,
even a reduction of the number of stars per magnitude bin by a factor of two
leads to
no {\it significantly} different distribution, and the hypothesis that
binaries cannot account for the difference cannot be ruled out with any
reasonable confidence. Henry \&
McCarthy (1990) arrive at the opposite conclusion to Stobie et al.
(1989) using a similar stellar sample.

The lesson to be learned here is that
conclusions with useful confidence are not possible if the investigation is
restricted to the nearby
sample only. Extension of the analysis to contain photographic star count data,
and using extensive numerical modeling, however, leads to statistically
significant conclusions (Kroupa et al. 1991, 1993).

\nobreak\vskip 10pt\nobreak
\noindent{\bf 3.2 They do matter!}
\nobreak\vskip 10pt\nobreak
\noindent
A detailed and consistent model of star count data is studied by Kroupa et al.
(1993). They construct physical models of all contributions to cosmic scatter
and measurement uncertainties in photometric and trigonometric parallax and
vary the proportion of unresolved binary
systems which they assume have component masses paired randomly from one
stellar mass function (i.e. an uncorrelated mass ratio distribution). The
resulting non-Gaussian
model cosmic scatter agrees with the observed value and is a function of
absolute magnitude. The spatial distribution of stars is assumed to follow an
exponential density law perpendicular to the Galactic plane. Realistic
modelling of the spatial distribution of stars is important in their analysis
because raw
star count data (i.e. not corrected for Malmquist bias nor a vertical density
gradient) are modelled.

Kroupa et al. (1993) show that
(i) binary systems lead to a significant
underestimation of faint star number densities in photographic surveys even if
only 50~per~cent of stars on the photographic plates are unresolved binaries
(fig.~21 in Kroupa et al. 1993) and
(ii) one single stellar mass function can `unify' both the nearby
and photographic star counts. They find that the initial mass function can be
approximated conveniently by the KTG($\alpha_1$) mass function with
$0.70<\alpha_1<1.85$ (equation~4 below).

\nobreak\vskip 10pt\nobreak
\noindent{\bf 3.3 Why do the conclusions of Reid (1991) and Kroupa et al.
(1993) differ?}
\nobreak\vskip 10pt\nobreak
\noindent
The models (A, B, C, E, F, G, H, I, L, M, N, O) computed by
Reid (1991) which have binaries of which either half or all have
equal-mass components are {\it not} consistent with observational
constraints:

The most thoroughly determined mass-ratio distributions by Duquennoy \& Mayor
(1991) and Mazeh et al. (1992) are for long- and short-period G-dwarf binary
systems, respectively. The long-period ($P>10^3$~days) systems mostly have
low-mass
companions, whereas the short-period systems ($P<10^3$~days) {\it may} show
some weak bias towards equal mass systems. However, of all G-dwarf binaries,
only about 10~per cent have $P<10^3$~days, of which only about 25~per cent have
a mass ratio $q>0.8$ (see also figs.~1 and~2 in Kroupa 1995b), where
$q={m_2\over m_1}\le1$,
where $m_1$ and $m_2$ are the masses of the primary ($m_1\approx1\,M_\odot$)
and scondary component, respectively. About 8~per
cent of all long-period G-dwarf binaries have $q>0.8$, so that of all G-dwarf
binaries only about 10~per cent have $q>0.8$, and the rest have $q<0.8$ with
about
55~per cent of all G-dwarf binaries having $q<0.4$. Similar conclusions are
arrived at from the preliminary results of the extensive radial velocity study
of K-dwarf binaries (Mayor et al. 1992). For M-dwarf binaries Fischer \& Marcy
(1992) conclude that the mass function of companion masses shows
no significant bias towards $q>0.8$.

The assumption made by
Reid (1991) in his models A, B, C, E, F, G, H, I, L, M, N and O that
half or all binaries have equal component masses
is thus inconsistent with the observational constraints.
We therefore concentrate only on those remaining models which are most
consistent with
observational constraints. These are his models D (50~per cent wide binaries),
J and K (100~per cent wide binaries).
Models D and J are based on a flat
model single star luminosity function, and model K is based on a model single
star luminosity function which has a maximum at about $M_{\rm V}\approx12$.

Concerning his models D and J,
we stress that the nearby star-count data {\it do not} require the nearby LF to
be flat
(figs.~1 and~3 in Kroupa 1995a).
The single star, single metallicity and single age luminosity function is
not likely to be flat for
$M_{\rm V}>12$ because changes in the first derivative of the
mass--absolute magnitude relation imply pronounced universal
structure at $M_{\rm V}\approx7$ (the `H$^-$ plateau') and at $M_{\rm
V}\approx12$ (the `H$_2$--convection peak') (Kroupa et al. 1990, 1993).
It is necessary to note
that while Malmquist-type bias does not significantly affect the stellar space
densities in the nearby luminosity function, its shape is smoothed because of
uncertainties in parallax measurements and the spread in metallicities (an age
spread does not significantly contribute to cosmic scatter for low-mass stars
with $V-I>2.5$ -- see fig.~4 in Kroupa et al. 1993). Comparison
of figs.~1 and~20 in Kroupa et al. (1993) demonstrates that the measured nearby
luminosity function {\it appears} much flatter than the true single
metallicity, single age luminosity function actually may be.
Reid's figs.~9 `D'
and `J' show that the model photometric luminosity functions do not decay as
steeply
with increasing $M_{\rm V}>12$ as his adopted observed photometric luminosity
function. We
take this as evidence that the single star luminosity function cannot be flat.

Concerning his model K, we find acceptable agreement in his fig.~9 `K'
between model photometric luminosity
function and adopted observed photometric luminosity function.
This model most closely resembles the
models found by Kroupa et al. (1993) to be consistent with the nearby {\it and}
photographic star count data.

An important reason why Kroupa et al. (1993) find that the nearby and
photometric
luminosity functions are manifestations of a single Galactic field stellar mass
function whereas Reid (1991) does not is because Kroupa et al. model a
realistic population of stellar systems and apply on this model population the
same
observational criteria as are used in the definition of the observed nearby and
photographic stellar samples. Reid's models lack the non-Gaussian nature of
cosmic scatter and its dependence on absolute magnitude. Contrary to his
assertion the reduction
of stellar number densities at distances between 100--300~pc from the Galactic
plane has to be taken into account when modelling raw star count data.
Consider a maximal case and a photographic survey perpendicular to the Galactic
plane with a photometric distance limit of $d_{\rm p}=130\,$pc (Stobie et al.
1989): An unresolved binary system with equal mass components will be included
if its distance is $\sqrt2\,d_{\rm p}=184\,$pc (Kroupa et al. 1991). If in
addition the combined absolute magnitude is $3\,\sigma_{\rm tot}=1.5\,$mag too
bright for its colour then this system will be included in the survey if its
true distance is as large as 367~pc. At this distance the stellar number
density has fallen to 0.3~of its value at the Galactic midplane assuming an
exponential Galactic disk scale height $h=300\,$pc. Clearly, a uniform spatial
distribution of stars to such distances is not an adequate approximation.
Taking account of this reduction of stellar number density with distance from
the Galactic midplane allows Kroupa et al. (1993) to account for the smaller
number densities in the observed photometric luminosity function (their
fig.~20).

Summarising, we interpret Reid's modelling as
follows: he investigates a region of parameter space not considered by Kroupa
et al. (1993) by studying other than uncorrelated mass ratio distributions in
the binary star model populations. His computational results lead to rejection
of the hypothesis that binary stars explain the difference between the nearby
and photometric luminosity functions, {\it unless} the mass ratio distribution
is approximately uncorrelated. This result is consistent
with the observed mass ratio distributions of G, K and M dwarf binaries. He
finds that binaries {\it can} account for the difference if the
single star single metallicity and age luminosity function has a maximum at
$M_{\rm V}\approx12$ (his model K). This finding verifies that of Kroupa et al.
(1991).
Kroupa et al. (1993), on the other hand, scan mass function power-law index,
$\alpha_1$, {\it and} disc scale height, $h$, space
under the assumption of an
uncorrelated mass ratio distribution (a good first approximation) which
consistently models the complex interdependence of normalisation {\it and}
shape of the model Malmquist uncorrected photometric luminosity function. They
find binaries {\it can} account for the difference between the nearby and
photometric luminosity functions if $0.70<\alpha_1<1.85$.

The
assumption by Kroupa et al. (1993) that the component masses in binary systems
are uncorrelated (note that
this implies that the mass function of the secondaries is steeper than the mass
function of the primaries) is a reasonable first approximation. For example,
Tout (1991) shows that simple selection effects can account for the
apparent correlation of component masses seen in single- and double-lined
spectroscopic binaries. However, some correlation of component masses is
present. Kroupa \& Tout (1992) point out that the G-dwarf binary systems
(Duquennoy \& Mayor 1991) have too few low-mass companions with respect to the
KTG($\alpha_1$) mass function if $\alpha_1>1$ approximately.

\nobreak\vskip 10pt\nobreak
\noindent{\bf 4 A REALISTIC MODEL}
\nobreak\vskip 10pt\nobreak
\noindent
If it is assumed that most stars form in embedded clusters, as suggested by
Lada \& Lada (1991), then most stars must also form in binary systems.
Conversely, if it is assumed that most stars form in binary systems (see e.g.
Mathieu 1994 and references therein) then the dominant mode of star formation
must be clustered star formation. Both conclusions are based on the observed
dynamical properties of Galactic field stars (Kroupa 1995b).

The dynamical properties of a stellar population are the mass function, the
proportion of binary systems and their distribution of orbital elements. In
Kroupa (1995b) we assume that all stars form in binary systems which have
component masses picked at random from the KTG(1.3) mass function (equation~4
below).
This assumption and our adopted initial distribution of periods are consistent
with pre-main sequence binary star observations. Our N-body simulations in
Kroupa (1995b) of a range of initial stellar aggregate sizes shows that
in a stellar aggregate consisting initially of a few hundred
binaries with half mass radius of about 0.8~pc
the initial dynamical properties evolve to the stellar dynamical
properties observed in the Galactic field. This is shown in detail in Kroupa
(1995c), where we perform 20~N-body simulations of this `dominant mode
cluster'. Specifically, the ionisation of binaries in the stellar aggregate
leads to a final binary proportion of 48~per cent and
the observed mass-ratio distribution of G-dwarf
binaries is obtained thus solving the discrepancy noted by Kroupa \& Tout
(1992), and thus also correctly reproducing the deviation from an uncorrelated
mass-ratio distribution.

We need to test whether the depletion of the mass ratio distribution at small
values invalidates the results of Kroupa et al. (1993). Using the extensive
data from our N-body simulations in Kroupa (1995c) we compute the single star
luminosity function and the system luminosity function
(merging the components of a binary system to an unresolved
system and counting stars not in binary systems as separate systems)
in the photometric V-, I- and K-bands and in bolometric magnitudes and
tabulate these in Appendix~1.
We restrict stellar masses to lie in the range
$0.1-1.1\,M_\odot$.

Kroupa et al. (1993) tabulate a mass--$M_{\rm V}$ relation obtained
by combining theoretical {\it and} observational (compiled by Popper
1980) constraints. This relation
is an excellent fit to the new mass--luminosity data compiled by Henry \&
McCarthy (1993). This is shown by Kroupa \& Gilmore (1994) by comparing the
mass$-M_{\rm V}$ data of Henry \& McCarthy (1993) that are not listed by
Popper
(1980) with the model mass$-M_{\rm V}$ relation of Kroupa et
al. (1993). This model mass$-M_{\rm V}$ relation may
prove usefull in constraining opacity and mixing-length theory.

Transformation to the K-band is obtained using the
$V-K,M_{\rm K}$ relation from Henry \& McCarthy (1993) (their equation~1a) and
to the I-band using the $M_{\rm V},V-I$ relation from Stobie et al.
(1989) (their equation~1) which provides a good fit to the data of
Monet et al. (1992) (we note that a slight correction of this relation owing to
systematic bias from unresolved binary stars is discussed by Kroupa et al.
1993). To obtain bolometric absolute magnitudes we use equation~6b in Kroupa
(1995a).

We compare our model luminosity functions with
the observational data in Fig.~1.
Our model luminosity function for individual stars, $\Psi_{\rm mod,sing}$,
is a good representation of the nearby data in the V-band
and in bolometric magnitudes. If all stars are
paired at random to form binary systems we obtain
the initial system luminosity function, $\Psi_{\rm mod,sys}(t=0)$ (`initial'
is to be understood here only w.r.t. pairing of component masses -- we do not
model pre-main sequence stellar evolution). After the dominant mode cluster
has disintegrated,
$\Psi_{\rm mod,sys}$, becomes the model system
luminosity function of the Galactic field, $\Psi_{\rm mod,sys}(t=1\,{\rm
Gyr})$. Our model system luminosity function is in good agreement with the
photometric luminosity function in the V-band. Agreement in bolometric
magnitudes is somewhat worse.

The residual discrepancies between $\Psi_{\rm mod,sys}(t=1\,{\rm Gyr})$ and
$\overline{\Psi}_{\rm phot}$  can partly be attributed to
remaining inadequacies of the model (mostly residual inaccuracies in the
mass--luminosity relation and the mass function -- it is probably not {\it
exactly} a power law below 0.5$\,M_\odot$). We do not model triple, quadruple,
etc. stellar systems. While the exact proportion of higher order systems is
unknown, it may be as large as 10--20~per cent (Abt \& Levy 1976, Duquennoy \&
Mayor 1991) of all binary systems. Taking account of these will cause
additional depression of the model system luminosity function bringing it into
better agreement with the photometric luminosity function at faint magnitudes.
Better agreement is achieved in the
V-band (upper panel in Fig.~1) than in bolometric magnitudes (lower panel in
Fig.~1) which is not surprising because our model was `fine-tuned' in the
V-band, and because the transformation to bolometric magnitudes
differs for the model data and the observational data. We care to point out
that bolometric corrections remain rather uncertain for low-mass stars. A
probably not negligible source of systematic uncertainty is treatment of
Malmquist
bias in the observed photometric luminosity functions. The best-estimate
Malmquist corrected photometric luminosity function was obtained from
Malmquist uncorrected photometric luminosity functions
(Kroupa 1995a) with the assumption that cosmic scatter is Gaussian
and independent of absolute magnitude, both of which are reasonable first
assumptions but not strictly true (Kroupa et al.
1993). Also, at $M_{\rm V}>15$, $M_{\rm Bol}>11$ photometric luminosity
functions may underestimate stellar number densities
because the $1/V_{\rm max}$ method
becomes unreliable if the underlying number of stars is too small (Jahreiss
1994 and references therein).
Furthermore, photometric calibration of photographic plates is non trivial and
uncertain for faint, very red stars. Given these considerations we warn
against
an over-interpretation of the residual discrepancies between model and data.

We do not exclude the possible enhancement at about $M_{\rm V}>16$ in stellar
number densities close to the Galactic midplane (Kirkpatrick et al. 1994).

Fig.~1 thus demonstrates that the discrepancy between the nearby and
photometric luminosity functions can be resolved by identifying the former with
the single star luminosity function (smoothed by the metallicity and
age dispersion
and parallax and photometric measurement errors), and the latter with the
system luminosity function. No additional
hypothesis needs to be postulated to account for the discrepancy between the
nearby and photometric luminosity functions.

\bigskip
\bigbreak
\noindent{\bf 5 THE STELLAR MASS FUNCTION}
\nobreak\vskip 10pt\nobreak
\noindent{\bf 5.1 Is there a significant maximum at about $0.3\,M_\odot$?}
\nobreak\vskip 10pt\nobreak
\noindent
Although the photometric luminosity function used by Tinney (1993),
$\Psi_{\rm TRM}^*$, significantly overestimates the stellar number densities
making it an incorrect estimator of the true stellar number density
distribution with
luminosity, its shape appears to be similar to our best-estimate
Malmquist corrected photometric
luminosity function, $\overline{\Psi}_{\rm phot}$ (fig.~4 in Kroupa 1995a).
Thus we take seriously the
suggestion by Tinney (1993) and Reid (1994) that the stellar mass function has
a maximum at $m\approx0.25\,M_\odot$, and we
investigate this matter more closely.

We note from our considerations in section~5 in Kroupa 1995a that
Tinney's survey contains nearly a factor two more stars at $M_{\rm Bol}=9$
($M_{\rm V}\approx10.5$),
than the nearby star sample. The mass function estimated by Scalo (1986)
assumes the stellar number density in the solar neighbourhood at this
magnitude. We would thus expect the mass functions which Tinney (1993) derives
from his luminosity function to lie above Scalo's mass function at
$m\approx0.45\,M_\odot$ by this same factor (the slope of the mass--$M_{\rm
V}$ relation is well defined at $M_{\rm V}\approx10.5$, see fig.~2 in Kroupa et
al. 1993). However, we find in all panels of fig.~7
of Tinney (1993) that the ``Scalo function seems to overpredict the number of
stars in the solar neighbourhood~.~.~. by about a factor of~2''.
This results from an incorrect scaling of the Scalo mass function in his fig.~7
a--g (Tinney, private communication).

We focus attention on fig.~7g of Tinney
(1993) and consider the mass--luminosity relation from Smith
(1983) used by Tinney in fig.~7g as representing the empirical relation, adopt
$M_{{\rm Bol},\odot}=4.72$, and obtain

$${\rm log}_{10}m=1.119-0.177\,M_{\rm Bol}, \eqno (1)$$

\noindent where $m$ is the
stellar mass in solar units. We follow Tinney (1993) and restrict this
relation to $0.2\,M_\odot<m<0.4\,M_\odot$. Thus,
$dM_{\rm Bol}/dm=-5.65/(m\,{\rm ln}10)$, and
the mass function in units of number of stars per pc$^3$ per solar mass is

$$\xi_{\rm Tin}(m)=2.454 {\Psi_{\rm TRM}^*(M_{\rm Bol})\over m}, \eqno (2)$$

\noindent which we tabulate in Table~1.

It is important to emphasise that equation~1 is not a valid representation of
the mass--luminosity data because it fails to account for the changes in
stellar structure and photometric properties due to the onset of full
convection and association
of H$_2$ at about $0.3\,M_\odot$. Structure in the mass function is inversely
proportional to the first derivative of the mass--absolute magnitude relation
which has a minimum at $M_{\rm V}\approx7$ and a pronounced maximum at $M_{\rm
V}\approx12$ (fig.~2 in Kroupa et al. 1993). Although theory
requires these features, Kroupa et al. 1990 and
recently D'Antona \& Mazzitelli
(1994) warn against blind use of theoretical
mass--luminosity relations, especially for stellar masses below $0.6\,M_\odot$,
because mixing-length theory and low temperature opacities are too uncertain.
The derivative of the theoretical mass--luminosity relation is thus
mostly useful for identification of structure and of the relevant physics, but
cannot be used to obtain detailed quantitative results. These
considerations should make it clear that any structure seen in the mass
function which is derived from any mass--luminosity relation is
very questionable.

In Table~1 Column~1 lists the absolute bolometric luminosity and Column~2 the
corresponding stellar mass obtained from equation~5. Column~3 contains the
photometric luminosity function used by Tinney (1993, his table~4), and
Column~4 tabulates the resulting mass function (equation~8). We also add a
short segment of the
mass function derived by Scalo (1986, his table~IV) by listing the absolute
visual magnitude (Column~5), the stellar mass (Column~6) and the Scalo mass
function which is obtained from his table~IV by evaluating

$$\xi_{\rm Sc}(m) = {\phi_{ms}({\rm log}_{10}m) \over 2h\,m\,{\rm ln}10},
\eqno (3)$$

\noindent where $\phi_{ms}$ is the number of stars per pc$^2$ per
$({\rm log}_{10}m)^{-1}$. Following Scalo the vertical Galactic disk scale
height is
$h=325$~pc (note that Tinney appears to adopt $h=350$~pc but references Scalo).

In Fig.~2 we plot $\xi_{\rm Tin}(m)$ and $\xi_{\rm Sc}(m)$.
In our figure the Scalo mass function
lies below the mass function derived from Tinney's luminosity function as
expected.

The mass function, $\xi_{\rm Tin}(m)$, which we obtain from Tinney's luminosity
function, has a maximum at a mass of approximately $0.23\,M_\odot$.
If we were to extend the mass--luminosity relation (equation~1) to
$m<0.2\,M_\odot$ then the mass function would decrease further, making the
maximum more apparent. Rather than recomputing $\xi_{\rm Tin}$ for all values
of $m$ we are satisfied that in his fig.~7g Tinney correctly plots his mass
function (compare with Table~1).  Tinney
(1993) also finds the maximum for all theoretical mass--luminosity relations
that he considers, and he and Reid (1994)
conclude that the mass function for low mass stars is not a power law but
shows significant structure with a maximum at $0.25\,M_\odot$.

Reid (1994) argues that this ``turnover in number density
is real - the only surveys that find a continuously increasing mass function
are those deriving masses from $M_{\rm V}$, a notoriously inaccurate mass
indicator
for very low mas stars''. We strictly contradict this statement for stellar
masses $m>0.1\,M_\odot$. While it is true that precise estimation
of the true luminosity of low mass stars is difficult in the V-band because
most of the radiation energy is emitted in the near infrared, the accuracy of
V-band
photometry cannot be questioned and provides very reliable estimation of the
true luminosity of stars down to $M_{\rm V}=17$. This is easily inferred from
fig.~10 of Monet et al. (1992) which shows a well defined main sequence
for $M_{\rm V}\le17$.
Furthermore, mass estimation using the V-band is just as reliable as any other
photometric band, as can be verified by consulting the mass-luminosity data
compiled in various photometric bands by Henry \& McCarthy (1993). Their data
show that for $0.08\,M_\odot<m<0.18\,M_\odot$ the dispersion in log$_{10}m$ for
the mass--$M_{\rm V}$ relation fit is 0.060, for the mass--$M_{\rm J}$ relation
fit it is 0.066, for the mass--$M_{\rm H}$ relation fit it is
0.054 and for the mass--$M_{\rm K}$ relation fit it is 0.067 (Henry \& McCarthy
1993). Kroupa \& Gilmore
(1994) compare the mass-$M_{\rm V}$ relation derived by Kroupa et al. (1993)
using V-band photometry with these data (not listed by Popper 1980), which were
derived at infrared
wavelengths. The agreement is very good. For $m<0.1\,M_\odot$ {\it any}
photometric band provides a poor mass estimator because of the long contraction
time scales for such stars (see e.g. D'Antona \& Mazzitelli 1994).
In any case, a rising mass function was also found by Henry \& McCarthy (1990)
based on $M_{\rm K}$.

In fig.~7 of Tinney (1993) only two panels show a difference between the
alleged peak at about $0.22\,M_\odot$ and the minimum at about $0.12\,M_\odot$
amounting to a factor of approximately 2.1 (fig.~7c and 7d). Both mass
functions
were derived using the mass--luminosity relation model D of Burrows, Hubbard \&
Lunine (1989) evaluated at two ages. Figs~7a and 7b, based on their model B
mass--luminosity relation at two ages, show a difference between maximum and
minimum of a factor of 1.2, and in figs~7e and 7f (based on mass--luminosity
relations from D'Antonna \& Mazzitelli 1986) the mass function merely flattens
in the range $0.15\,M_\odot<m<0.3\,M_\odot$. The mass function in Tinney's
fig.~7g is based on the linear mass--luminosity relation, and shows a
difference between peak and minimum of a factor of 1.6.

Thus, of the seven mass functions Tinney computes from his uncorrected
luminosity function, only three show structure which amounts to a difference
between alleged peak and minimum of more than 50~per cent. One of these is
based on an inadequate linear mass--luminosity relation, and the other two are
based on one theoretical mass--luminosity relation. The remaining four mass
functions show structure amounting to less than a 20~per cent difference
between peak and minimum and are based on only two theoretical
mass--luminosity relations. It is important to bear in mind that the
theoretical
mass--luminosity relations used by Tinney (1993) are based on the same opacity
tables and use similar equations of state, so that the physical parameter space
scanned by Tinney (1993) is limited.

Having discussed the limited evidence for structure in the stellar mass
function we must, however, keep in mind that
Tinney (1993) computes the mass
functions using the Malmquist uncorrected bolometric photometric luminosity
function. Correcting for Malmquist bias would enhance the alleged maximum in
the mass function.

The observed stellar luminosity function has the kind of structure we see in
the first
derivative of the theoretical mass--absolute magnitude relation (Kroupa et al.
1990, 1993). By Occam's Razor we thus argue that the plateau at $M_{\rm
V}\approx7$ and the maximum at $M_{\rm V}\approx12$ in the luminosity function
reflect the changes in the slope of the mass--lumninosity relation. While this
does not disprove a
maximum in the stellar mass function, it would appear a remarkable coincidence
if main sequence stellar structure and star formation both lead to the same
structure in the stellar luminosity function. Our assumption leads to a model
consistent with star count data and can be verified observationally because
luminosity functions of {\it all} stellar populations must show this same
structure, independent of the stellar mass function (Kroupa et al. 1993).

In view of the uncertainties in stellar physics
below $0.6\,M_\odot$ (Kroupa et al. 1990, D'Antonna \& Mazzitelli 1994), the
presence of a plateau and maximum in the luminosity function at just the
correct absolute magnitudes where theory suggests the same features in the
slope of the mass--absolute magnitude relation, the
lack by Tinney to account for unresolved binary stars in his sample
(which can depress the
photometric luminosity function at $M_{\rm V}\approx15$ by a factor of more
than~2.5 -- see fig.~21 in Kroupa et al. 1993) and the
discrepancy in stellar number densities in his luminosity function (corrected
and uncorrected)  at $M_{\rm Bol}<9.5$ (Kroupa 1995a), we argue
that the structure in the mass function found by Tinney is not significant. All
these effects distort the shape of the luminosity function making {\it any}
structure seen in the  mass function derived from a photometric luminosity
function using the (incorrect) `standard technique' [(photometric luminosity
function)=(mass function)/(slope of mass--absolute magnitude relation)] very
questionable indeed.

Similar reservations
apply to the result obtained by Strom, Strom \& Merrill (1993) that the
mass function for the young stellar population associated with the L1641
molecular cloud (at a distance of~480~pc) peaks near $0.3\,M_\odot$. They base
their conclusion on a theoretical mass--luminosity relation corrected for
pre-main sequence brightening, which significantly adds to the
uncertainties discussed above. Furthermore, at a distance of 480~pc virtually
all binary systems remain unresolved. Strom et al. (1993) neglect to correct
for these which can lead to serious underestimation of the number of faint
stars
because the proportion of pre-main sequence binary systems must be assumed to
be higher than in the Galactic field (see Mathieu 1994 and references
therein, Kroupa 1995b).

Finally, we
emphasise that in their section~9 Kroupa et al. (1993) explicitly test a model
of star count data based on a mass function with a maximum at about
0.3$\,M_\odot$ and find poor agreement with observed star counts. It is also
worth noting that Kroupa et al. (1991) mention that they experienced difficulty
unifying both the nearby and photometric luminosity functions using other than
a power law mass function below $0.5\,M_\odot$.

\nobreak\vskip 10pt\nobreak
\noindent{\bf 5.2 No signifacant maximum!}
\nobreak\vskip 10pt\nobreak
\noindent
Consistent treatment of photographic {\it and} nearby star count data
{\it and} of the mass--luminosity relation
with detailed modelling of cosmic scatter (see also Section~3.2) provides good
solutions to raw star count data if the
number of stars at birth,
$\xi(m)\,dm$, in the mass interval $m$ to $m+dm$ (in solar units) is
approximated by

$$\xi (m) = k_\xi \cases{{\displaystyle 0.5^{\alpha_1}\,m^{-\alpha_1}},
&if $0.08 \le m < 0.5$; \cr
{\displaystyle 0.5^{2.2}\,m^{-2.2}},
&if $0.5 \le m < 1.0$; \cr
{\displaystyle 0.5^{2.2}\,m^{-2.7}},
&if $1.0 \le m < \infty$, \cr}\eqno (4)$$

\noindent where $k_\xi=0.0873$ stars pc$^{-3}$ $M_\odot^{-1}$ if
$\alpha_1=1.3$ scales to the observed stellar number density. We
refer to equation~4 as the KTG($\alpha_1$) initial mass function. Allowing for
unresolved binary stars and a possible metallicity gradiant perpendicular to
the Galactic disc Kroupa et al. (1993) constrain $\alpha_1$ to lie in the
95~per cent confidence interval 0.70--1.85.

The nearby data place poorer constraints on the stellar
mass function.
Considering the nearby star count
data alone and applying the Kolmogorov-Smirnov test on a model nearby sample
which accounts for trigonometric parallax and photometric
measurement uncertainties and a detailed physical model of the dispersion in
stellar luminosities due to a spread in metallicities and ages, the
95~per cent confidence interval for $\alpha_1$ is 0.6--2.4 (fig.~16 in Kroupa
et al. 1993).
Fitting a power-law mass function to the nearby luminosity
function Haywood (1994) finds $1.3<\alpha<1.9$ (confidence interval is
unspecified) for $m<0.35\,M_\odot$, but notes
that a change in $\alpha$ at about $0.35\,M_\odot$is not required by the data.
His result is based on a theoretical mass--luminosity relation and does not
account for the dispersion in metallicities and ages, nor for measurement
errors. Using the mass$-M_{\rm V}$ relation of Kroupa
et al. (1993) and the KTG($\alpha_1$) mass function without modelling the
metallicity and age dispersion and measurement errors
we compute model single star luminosity functions, which we scale to
the nearby
luminosity function at $5.5\le M_{\rm V}\le8.5$ (0.0106~stars pc$^{-3}$,
table~2 in Kroupa 1995a).
A $\chi^2$ test on the nearby luminosity function at $12.5\le
M_{\rm V}\le16.5$ (observed value: $0.051\pm0.011$~stars pc$^{-3}$) constrains
$\alpha_1$ to lie in the range 0.66--1.44 with 95~per cent confidence. The
corresponding ideal single star model luminosity functions are plotted in
Fig.~3.

Comparison with the above result by Kroupa et al.
(1993) illustrates that
different statistical tests together with proper modelling of cosmic scatter
and measurement errors can yield different results. It is always advisable to
use as many different
tests as possible, and we note that consistent modelling of cosmic scatter
and measurement errors together with the Kolmogorov-Smirnov test yields more
conservative (i.e. wieder)
bounds on $\alpha_1$ than the more constraining $\chi^2$ test. We stress
however, that our fitting here ($0.66<\alpha_1<1.44$), that of Haywood (1994)
($1.3<\alpha<1.9$), and Kroupa et al. (1993) ($0.70<\alpha_1<1.85$) are all
consistent with $\alpha_1\approx1.3$.

We emphasise that in our model the maximum in the
photometric luminosity function at $M_{\rm V}\approx12$, $M_{\rm
Bol}\approx9.7$ is reproduced
despite the absence of any such structure in the power-law mass function.
It results from a maximum in the first derivative of the mass--absolute
magnitude relation because of changes in the stellar constitution at about
$0.33\,M_\odot$ owing to the onset of association of H$_2$ and the onset of a
fully convective interior. This feature in the stellar luminosity function is
thus universal, i.e. independent of the stellar mass function (Kroupa et al.
1993).

The KTG(1.3) mass function of the Galactic field star population
is plotted as the thick solid line in Fig.~2.
It has the same stellar number densities per mass interval as Scalo's mass
function for $m>0.25\,M_\odot$ and the latter does not overpredict the number
of stars as suggested by Tinney (1993). At lower masses, $\xi_{\rm Sc}(m)$
underestimates the number densities of stars because Scalo (1986) does not
allow for unresolved binaries in his adopted luminosity function. Also,
$\xi_{\rm Sc}$ follows the shape of the luminosity function adopted by Scalo
(1986) because he uses an empirical, approximately linear, log(mass)--$M_{\rm
V}$ relation. The
KTG($\alpha_1$) mass function is plotted and
compared with $\xi_{\rm Sc}(m)$ for all stellar masses in fig.~22 of Kroupa et
al. (1993).

It is of interest to note that our inverse dynamical population synthesis
(Kroupa 1995b) suggests that there may be a relationship between $\alpha_1$ and
the star formation mode. The dynamical properties of Galactic field systems are
best reproduced if most stars form in the dominant mode cluster (see
Section~4) {\it and} if $\alpha_1\approx1.3$.

\bigskip
\bigbreak
\noindent{\bf 6 CONCLUSIONS}
\nobreak\vskip 10pt\nobreak
\noindent
The highly significant difference
between the nearby and Malmquist corrected photometric luminosity functions
(Kroupa 1995a) can
be resolved naturally if the nearby luminosity function is identified with the
single star
luminosity function, and if the Malmquist corrected photometric
luminosity function is identified
with the system luminosity function (Fig.~1).
To this end we introduce in Section~4 a model of the dynamical properties of
stellar systems which is consistent with all presently available observational
constraints. The model luminosity functions are tabulated in the photometric
V-, I- and K-bands and in bolometric magntitudes in Appendix~1.

It is
important to emphasise that our model system luminosity function is a
reasonable fit to the Malmquist corrected photometric luminosity function
because
cosmic scatter and photometric errors have been removed from the photometric
luminosity function. These lead to Malmquist bias which, among other
effects, broadens or smooths structure in the photometric luminosity function.
The nearby luminosity function, however, has not been corrected for the
metallicity and age spread nor for photometric and trigonometric parallax
measurement
errors. The pronounced `H$_2$--convection maximum' at $M_{\rm V}\approx12$,
which we expect on the basis of our ideal (i.e. single metallicity and age,
no measurement errors) single-star model is thus smoothed out and swamped by
the large statistical uncertainties.

Our result that binaries account for the difference
agrees with the conclusions
by Kroupa et al. (1993) who modelled raw star count data. No additional
hypothesis to account for the difference is required.

The alternative
hypothesis of a local overdensity of faint stars does not satisfactorily
account for the difference because a
highly improbable and special stellar constellation needs to be postulated
(Section~2).

The objections by Reid (1991) that
binaries cannot account for the difference are valid only for the particular
(`favorite') model parameters he considers. These, however, are inconsistent
with observational mass-ratio distributions and with evidence from
theoretical
stellar models that the single star luminosity function is
likely to have a pronounced maximum.  Rather than concluding from his
investigation that unresolved binary systems cannot account for the
difference,
the correct interpretation of his modelling is that it shows that his
`favorite' models are inconsistent with star count data.

The stellar mass function, which `unifies' both the nearby {\it and}
photometric luminosity functions, can be approximated by the
KTG(1.3) mass function (equation 4).

The nearby star count data alone constrain the stellar mass function
poorly. To significantly improve the statistical uncertainties of the
single star
luminosity function the stellar sample in which all binaries are resolved and
which is complete to the faintest stellar luminosities has to be increased by
{\it at least} an order of magnitude. This requires significant future effort
in large-scale trigonometric parallax surveys to identify all faint stars in
close proximity (e.g. within about 5~pc in the southern hemisphere
and/or to distances larger than the 5.2~pc distance limit), and infrared
speckle observations to identify all binary systems in the parallax survey.

There is no evidence for structure in the stellar mass function for low mass
stars beyond the change in $\alpha_i$ at $m=0.5\,M_\odot$ (equation~4). A
maximum in the mass
function at a mass of approximately $0.25\,M_\odot$ is found by researchers who
(i) restrict their analysis to the photometric luminosity function (or a
stellar luminosity function for a population of stars at sufficiently large
distance so that the observational apparatus cannot resolve the majority of
binary systems),
(ii) do not correct for unresolved binaries, and (iii) use inadequate
mass--luminosity relations (e.g. a linear log(mass)--absolute magnitude
relation). In their section~9 Kroupa et al.
(1993) show that
a mass function with a maximum at about $0.3\,M_\odot$ is a poor fit to
the nearby {\it and} photometric luminosity functions.

Robust evidence in support of the proposition that there
exists structure in the mass function beyond the flattening at $0.5\,M_\odot$
does not presently exist. The extensive star count data obtained by Tinney et
al. (1993) may shed light on this problem, but only after
Malmquist bias has been taken care of very carefully, after the
reason for the significant
deviation of stellar number density at $M_{\rm bol}<9.8$ in their photometric
luminosity function from the density in the other photometric luminosity
functions (figs.~4 and~5 in Kroupa 1995a) has been resolved,
and after unresolved
binary systems have been accounted for. To this end a detailed study of cosmic
scatter for faint stars must be performed in the V-, R-, and I-bands.
However, even then the residual uncertainties
in the mass--luminosity relation will make conclusions concerning structure in
the mass function beyond a flattening questionable.

Because cosmic scatter and the effects of binaries lead to a non-Gaussian
dispersion of stellar absolute magnitudes which is also absolute magnitude
dependent, we strongly urge researchers studying the
stellar mass function to proceed along the lines of Kroupa et al. (1993), who
model cosmic scatter and raw star count data (which are filtered to exclude
galaxies, white dwarfs, and giant stars),
rather than using the incorrect `standard technique' of deriving a mass
function directly from the photometric (or low spatial resolution) luminosity
function.

\bigskip
\bigbreak
\par\noindent{\bf ACKNOWLEDGMENTS}
\nobreak
I thank T. Henry and T. Simon for carefully reading the manuscript and many
suggestions which improved the presentation.
I also thank H. Bernstein, G. Gilmore, N. Reid, C. Tinney and C. A. Tout
for useful and much appreciated comments.

\vfill\eject

\bigbreak
\vskip 3mm
\bigbreak

\noindent
{\bf Table 1.} The stellar mass function from $\Psi_{\rm TRM}^*(M_{\rm
Bol})$ and from Scalo (1986).

\nobreak
\vskip 1mm
\nobreak
{\hsize 13 cm \settabs 11 \columns

\+$M_{\rm Bol}$ &$m$ &$\Psi_{\rm TRM}^*(M_{\rm Bol})$ &&&$\xi_{\rm Tin}(m)$
&&&$M_{\rm V}$ &$m$ &$\xi_{\rm Sc}(m)$ \cr

\+  &$M_\odot$ &$\times10^{-3}{\rm pc}^{-3}{\rm mag}^{-1}$
&&&${\rm pc}^{-3}M_\odot^{-1}$
&&&            &$M_\odot$ &${\rm pc}^{-3}M_\odot^{-1}$ \cr

\+9.00 &0.336 &27.007 &&&0.197 &&&10 &0.447 &0.101\cr
\+9.25 &0.303 &25.946 &&&0.210 &&&11 &0.363 &0.140\cr
\+9.50 &0.274 &27.622 &&&0.247 &&&12 &0.288 &0.176\cr
\+9.75 &0.247 &32.074 &&&0.318 &&&13 &0.224 &0.226\cr
\+10.00 &0.223 &30.879 &&&0.340 &&&14 &0.178 &0.266\cr
\+10.25 &0.202 &24.944 &&&0.303 &&&15 &0.141 &0.211\cr
}
\bigbreak\vskip 3mm


\vfill\eject

\bigskip
\noindent{\bf REFERENCES}
\nobreak
\bigskip
\nex Abt, H. A., Levy, S. G., 1976, ApJS 30, 273
\nex Burrows, A., Hubbard, W. B., Lunine, J. I., 1989, ApJ 345, 939
\nex D'Antona, F., Mazzitelli, I., 1985, ApJ, 296, 502
\nex D'Antona, F., Mazzitelli, I., 1994, ApJS 90, 467
\nex Duquennoy, A., Mayor, M., 1991, A\&A, 248, 485
\nex Fischer, D. A., Marcy, G. W., 1992, ApJ 396, 178
\nex Haywood, M., 1994, A\&A 282, 444
\nex Henry, T. J., McCarthy, D. W., 1990, ApJ 350, 334
\nex Henry, T. J., McCarthy, D. W., 1993, AJ 106, 773
\nex Jahreiss, H., 1994, Ap\&SS 217, 63
\nex Kirkpatrick, J. D., McGraw, J. T., Hess, T. R., Liebert, J., McCarthy, D.
     W., 1994, ApJS 94, 749
\nex Kroupa, P., 1995a, Are the Nearby and Photometric Stellar Luminosity
     Functions Different?, ApJ, in press
\nex Kroupa, P., 1995b, Inverse Dynamical Population Synthesis and Star
     Formation, in preparation
\nex Kroupa, P., 1995c, The Dynamical Properties of Stellar Systems in the
     Galactic Disc, in preparation
\nex Kroupa, P., Gilmore, G., 1994, MNRAS 269, 655
\nex Kroupa, P., Tout, C. A., 1992 MNRAS, 259, 223
\nex Kroupa, P. Tout, C. A., Gilmore, G., 1990, MNRAS 244, 76
\nex Kroupa, P. Tout, C. A., Gilmore, G., 1991, MNRAS 251, 293
\nex Kroupa, P., Tout, C. A., Gilmore, G., 1993, MNRAS 262, 545
\nex Lada, C. J., Lada, E. A., 1991, The Nature, Origin and Evolution of
     Embedded Star Clusters. In: James, K. (ed.), The Formation and Evolution
     of Star Clusters, ASP Conf. Series 13, 3
\nex Mathieu, R. D., 1994, ARA\&A, 32, 465
\nex Mayor, M, Duquennoy, A., Halbwachs, J.-L., Mermilliod, J.-C.,
     1992, CORAVEL Surveys to Study Binaries at Different Masses and Ages. In:
     McAlister, H. A., Hartkopf, W. I. (eds.) Complementary Approaches to
     Double and Multiple Star Research, Proc. IAU Coll. 135, ASP Conference
     Series, 32, p.73
\nex Mazeh, T., Goldberg, D., Duquennoy, A., Mayor, M., 1992, ApJ 401, 265
\nex Monet, D. G., Dahn, C. C., Vrba, F. J., Harris, H. C., Pier, J. R.,
     Luginbuhl, C. B., Ables, H. D., 1992, AJ, 103, 638
\nex Popper, D. M., 1980, ARA\&A 18, 115
\nex Reid, N., 1987, MNRAS 225, 873
\nex Reid, N., 1991, AJ 102, 1428
\nex Reid, N., 1994, Ap\&SS 217, 57
\nex Scalo, J. M., 1986, Fund. Cosmic Phys. 11, 1
\nex Smith, R. C., 1983, Observatory 103, 29
\nex Stobie, R. S., Ishida, K., Peacock, J. A., 1989, MNRAS 238, 709
\nex Strom, K. M., Strom, S. E., Merrill, K. M., 1993, ApJ 412, 233
\nex Tinney, C. G., 1993, ApJ 414, 279
\nex Tinney, C. G., Reid, N., Mould, J. R., 1993, ApJ 414, 254
\nex Tout, C. A., 1991, MNRAS 250, 701

\vfill\eject


\bigbreak
\vskip 3mm
\bigbreak

\hang{ {\bf APPENDIX 1: The Stellar Luminosity Function} (Section~4)}

\nobreak
\vskip 1mm
\nobreak
{\hsize 15 cm \settabs 7 \columns

\+& $M_{\rm V}$ &$\Psi_{\rm mod,sing}$ &$\delta \Psi$
&$\Psi_{\rm mod,sys}$ &$\delta \Psi$
&$\Psi_{\rm mod,sys}$ &$\delta \Psi$ \cr
\+& && &~~~~~~$t=0$ & &~~~~$t=1\,$Gyr \cr
\+\cr
\+&     1.0 &   0.00 &   0.00  &      0.00 &   0.00  &      0.00 &  0.00 \cr
\+&     1.5 &   0.05 &   0.05  &      0.05 &   0.05  &      0.10 &  0.07 \cr
\+&     2.0 &   0.30 &   0.12  &      0.30 &   0.12  &      0.35 &  0.13 \cr
\+&     2.5 &   0.25 &   0.11  &      0.25 &   0.11  &      0.50 &  0.16 \cr
\+&     3.0 &   0.15 &   0.08  &      0.15 &   0.08  &      0.45 &  0.15 \cr
\+&     3.5 &   0.25 &   0.11  &      0.80 &   0.20  &      0.65 &  0.18 \cr
\+&     4.0 &  1.10 &   0.24  &      0.75 &   0.19  &     1.20 &  0.25 \cr
\+&     4.5 &  3.40 &   0.42  &     3.45 &   0.42  &     3.35 &  0.41 \cr
\+&     5.0 &  5.40 &   0.53  &     4.95 &   0.51  &     4.45 &  0.48 \cr
\+&     5.5 &  3.25 &   0.41  &     2.40 &   0.35  &     2.65 &  0.37 \cr
\+&     6.0 &  5.20 &   0.52  &     5.35 &   0.53  &     4.75 &  0.50 \cr
\+&     6.5 &  5.40 &   0.53  &     6.85 &   0.60  &     5.45 &  0.53 \cr
\+&     7.0 &  9.30 &   0.69  &     5.00 &   0.51  &     6.65 &  0.59 \cr
\+&     7.5 &  3.95 &   0.45  &     3.25 &   0.41  &     3.80 &  0.44 \cr
\+&     8.0 &  2.80 &   0.38  &     2.30 &   0.34  &     2.50 &  0.36 \cr
\+&     8.5 &  8.45 &   0.66  &     8.65 &   0.67  &     8.20 &  0.65 \cr
\+&     9.0 &  7.35 &   0.62  &     6.80 &   0.59  &     6.95 &  0.60 \cr
\+&     9.5 & 12.75 &   0.81  &     9.55 &   0.70  &    10.55 &  0.74 \cr
\+&    10.0 & 10.95 &   0.75  &    12.00 &   0.79  &    11.90 &  0.79 \cr
\+&    10.5 & 19.95 &  1.02  &    19.85 &  1.02  &    19.85 & 1.02 \cr
\+&    11.0 & 25.35 &  1.15  &    21.20 &  1.05  &    22.90 & 1.09 \cr
\+&    11.5 & 34.80 &  1.35  &    26.60 &  1.18  &    28.15 & 1.21 \cr
\+&    12.0 & 43.35 &  1.51  &    21.85 &  1.07  &    29.55 & 1.24 \cr
\+&    12.5 & 41.70 &  1.48  &     8.85 &   0.68  &    18.95 &  0.99 \cr
\+&    13.0 & 27.75 &  1.20  &    10.15 &   0.73  &    15.10 &  0.89 \cr
\+&    13.5 & 23.40 &  1.10  &     9.15 &   0.69  &    13.55 &  0.84 \cr
\+&    14.0 & 24.45 &  1.13  &     3.85 &   0.45  &    10.80 &  0.75 \cr
\+&    14.5 & 19.80 &  1.02  &     2.10 &   0.33  &     8.55 &  0.67 \cr
\+&    15.0 & 17.70 &   0.96  &     3.40 &   0.42  &     9.05 &  0.69 \cr
\+&    15.5 & 23.75 &  1.11  &      0.00 &   0.00  &     9.10 &  0.69 \cr
\+&    16.0 & 11.50 &   0.77  &      0.00 &   0.00  &     4.50 &  0.48 \cr
\+&    16.5 &   0.00 &   0.00  &      0.00 &   0.00  &      0.00 &  0.00 \cr

\vfill\eject

\+& $M_{\rm K}$ &$\Psi_{\rm mod,sing}$ &$\delta \Psi$
&$\Psi_{\rm mod,sys}$ &$\delta \Psi$
&$\Psi_{\rm mod,sys}$ &$\delta \Psi$ \cr
\+& && &~~~~~~$t=0$ & &~~~~$t=1\,$Gyr \cr
\+\cr
\+&      0.0 &   0.00 &  0.00  &     0.00 &  0.00  &      0.00 &  0.00 \cr
\+&      0.5 &   0.00 &  0.00  &     0.00 &  0.00  &      0.00 &  0.00 \cr
\+&     1.0 &   0.00 &  0.00  &     0.00 &  0.00  &      0.05 &  0.05 \cr
\+&     1.5 &   0.20 &  0.10  &     0.25 &  0.11  &      0.45 &  0.15 \cr
\+&     2.0 &   0.55 &  0.17  &     0.55 &  0.17  &     1.00 &  0.22 \cr
\+&     2.5 &   0.25 &  0.11  &    1.00 &  0.22  &     1.00 &  0.22 \cr
\+&     3.0 &  4.00 &  0.45  &    4.95 &  0.51  &     4.70 &  0.49 \cr
\+&     3.5 &  5.95 &  0.55  &    7.65 &  0.63  &     6.15 &  0.56 \cr
\+&     4.0 & 17.75 &  0.96  &   14.10 &  0.86  &    14.00 &  0.85 \cr
\+&     4.5 & 11.20 &  0.76  &    8.85 &  0.68  &     9.95 &  0.72 \cr
\+&     5.0 & 11.55 &  0.77  &   14.85 &  0.88  &    14.15 &  0.86 \cr
\+&     5.5 & 20.05 & 1.02  &   18.50 &  0.98  &    18.80 &  0.99 \cr
\+&     6.0 & 23.80 & 1.11  &   27.20 & 1.19  &    25.40 & 1.15 \cr
\+&     6.5 & 36.80 & 1.39  &   41.15 & 1.47  &    40.70 & 1.46 \cr
\+&     7.0 & 57.20 & 1.73  &   29.15 & 1.23  &    38.70 & 1.42 \cr
\+&     7.5 & 61.80 & 1.80  &   18.30 &  0.98  &    31.15 & 1.28 \cr
\+&     8.0 & 50.40 & 1.62  &    9.70 &  0.71  &    22.20 & 1.08 \cr
\+&     8.5 & 43.80 & 1.51  &    3.65 &  0.43  &    18.05 &  0.97 \cr
\+&     9.0 & 48.50 & 1.59  &     0.00 &  0.00  &    18.05 &  0.97 \cr
\+&     9.5 &   0.00 &  0.00  &     0.00 &  0.00  &      0.00 &  0.00 \cr

\vskip 5mm

\+& $M_{\rm I}$ &$\Psi_{\rm mod,sing}$ &$\delta \Psi$
&$\Psi_{\rm mod,sys}$ &$\delta \Psi$
&$\Psi_{\rm mod,sys}$ &$\delta \Psi$ \cr
\+&& & &~~~~~~$t=0$ & &~~~~$t=1\,$Gyr \cr
\+\cr
\+&   0.5 &   0.00&   0.00 &    0.00&   0.00 &     0.00 &   0.00  \cr
\+&  1.0 &   0.00&   0.00 &    0.00&   0.00 &     0.05 &   0.05  \cr
\+&  1.5 &   0.20&   0.10 &    0.20&   0.10 &     0.15 &   0.08  \cr
\+&  2.0 &   0.25&   0.11 &    0.25&   0.11 &     0.60 &   0.17  \cr
\+&  2.5 &   0.30&   0.12 &    0.30&   0.12 &     0.60 &   0.17  \cr
\+&  3.0 &   0.25&   0.11 &    0.80&   0.20 &     0.70 &   0.19  \cr
\+&  3.5 &  2.05&   0.32 &   2.05&   0.32 &    2.35 &   0.35  \cr
\+&  4.0 &  4.60&   0.49 &   4.35&   0.47 &    4.30 &   0.47  \cr
\+&  4.5 &  5.55&   0.54 &   4.65&   0.49 &    4.35 &   0.47  \cr
\+&  5.0 &  6.20&   0.57 &   7.60&   0.63 &    6.20 &   0.57  \cr
\+&  5.5 & 11.10&   0.76 &   8.30&   0.66 &    8.65 &   0.67  \cr
\+&  6.0 &  6.45&   0.58 &   5.20&   0.52 &    5.80 &   0.55  \cr
\+&  6.5 &  5.85&   0.55 &   6.70&   0.59 &    6.50 &   0.58  \cr
\+&  7.0 & 11.95&   0.79 &  11.65&   0.78 &   11.95 &   0.79  \cr
\+&  7.5 & 16.60&   0.93 &  15.20&   0.89 &   15.35 &   0.89  \cr
\+&  8.0 & 18.25&   0.98 &  24.15&  1.12 &   21.75 &  1.06  \cr
\+&  8.5 & 36.05&  1.37 &  36.10&  1.37 &   37.30 &  1.40  \cr
\+&  9.0 & 52.30&  1.65 &  32.80&  1.31 &   39.15 &  1.43  \cr
\+&  9.5 & 63.75&  1.83 &  14.80&   0.88 &   29.20 &  1.23  \cr
\+& 10.0 & 40.40&  1.45 &  14.45&   0.87 &   22.45 &  1.08  \cr
\+& 10.5 & 31.85&  1.29 &   5.00&   0.51 &   13.65 &   0.84  \cr
\+& 11.0 & 29.35&  1.24 &   4.85&   0.50 &   13.90 &   0.85  \cr
\+& 11.5 & 31.45&  1.28 &    0.45&   0.15 &   12.05 &   0.79  \cr
\+& 12.0 & 19.05&  1.00 &    0.00&   0.00 &    7.50 &   0.62  \cr
\+& 12.5 &   0.00&   0.00 &    0.00&   0.00 &     0.00 &   0.00  \cr

\vfill\eject

\+& $M_{\rm bol}$ &$\Psi_{\rm mod,sing}$ &$\delta \Psi$
&$\Psi_{\rm mod,sys}$ &$\delta \Psi$
&$\Psi_{\rm mod,sys}$ &$\delta \Psi$ \cr
\+& && &~~~~~~$t=0$ & &~~~~$t=1\,$Gyr \cr
\+\cr
\+&   8.0&  15.20&    0.89& 14.90&    0.88&    15.50&    0.90 \cr
\+&   8.5&  17.20&    0.95& 25.95&   1.16&    21.55&   1.06 \cr
\+&   9.0&  33.65&   1.33& 29.55&   1.24&    32.05&   1.29 \cr
\+&   9.5&  40.80&   1.46& 40.50&   1.46&    39.50&   1.44 \cr
\+&  10.0&  61.05&   1.79& 27.65&   1.20&    37.55&   1.40 \cr
\+&  10.5&  45.00&   1.53& 21.15&   1.05&    27.15&   1.19 \cr
\+&  11.0&  42.95&   1.50& 10.20&    0.73&    19.75&   1.01 \cr
\+&  11.5&  41.00&   1.46&  8.15&    0.65&    19.10&   1.00 \cr
\+&  12.0&  45.05&   1.53&  3.80&    0.44&    18.10&    0.97 \cr

}
\bigbreak\vskip 3mm

\hang{
These data are a model of an equal metallicity and age population of stars and
binaries for which ideal photometry and distance information are available.
It is based on the mass--$M_{\rm V}$ relation tabulated by Kroupa et al. (1993)
and the KTG(1.3) mass function. We restrict stellar masses to lie in the range
$0.1-1.1\,M_\odot$. This model stellar population
results if the stars form
in aggregates that are dynamically equivalent to the dominant mode cluster
(Kroupa 1995c). The above tables list
the luminosity functions in the V-, K- and I-bands, and also the bolometric
luminosity functions. For each photometric band column~1 contains the absolute
magnitude. Columns~2 and~3 contain the luminosity function obtained if all
stars
are counted individually and the Poisson uncertainty, respectively. Columns~4
and~5 list
the initial system luminosity function (pre-main sequence brightening is not
included) and the uncertainty, respectively. This luminosity function is
obtained if all
stars are paired at random from the KTG(1.3) mass function giving a binary
proportion of unity and an uncorrelated mass ratio distribution with slight
adjustment for `feeding' during `pre-main sequence eigenevolution'
(for details see Kroupa 1995c). Columns~6 and~7 list the final system
luminosity function and the uncertainty, respectively. This luminosity function
corresponds to a population of Galactic field systems of which 48~per cent are
binaries with orbital parameters as discussed in Section~4,
and has evolved from the luminosity function listed under Column~4 after
aggregate disintegration.
The luminosity functions and the uncertainties tabulated here are averaged from
$N_{\rm run}=20$ N-body simulations, each of an aggregate consisting of
200~binary systems
(table~1 in Kroupa 1995b). Thus $\Psi =
{1\over N_{\rm run}}\sum^{N_{\rm run}}_{i=1} \, N_{Mi}$ and
the uncertainty,
$\delta\Psi = \left[ {\sum^{N_{\rm run}}_{i=1}  \, N_{Mi}
                      \over (N_{\rm run}-1)} \right]^{1\over2}\,
N_{\rm run}^{-{1\over2}}$ (standard deviation of the mean),
where $N_{Mi}$ is the number of stars or systems in absolute magnitude
bin $M_i$ in simulation $i$.
}


\vfill\eject

\centerline{\bf Figure captions}
\smallskip

\noindent {\bf Figure~1.} Comparison of our ideal model (i.e.
single-metallicity and age and no measurement errors) Galactic field
luminosity
functions (Appendix~1) with observations in the photometric V-band (upper
panel) and in bolometric magnitudes (lower panel) (Section~4). The solid-line
histogram represents the observed
nearby stellar luminosity function, $\Psi_{\rm near}$, which is not corrected
for Malmquist-type bias (tables~2 and~8 in Kroupa 1995a) and which is smoothed
at the faint end as detailed in section~4 of Kroupa (1995a).
The filled cicles represent our best estimate Malmquist corrected photometric
luminosity function, ${\overline\Psi}_{\rm phot}$
(tables~2 and~8 in Kroupa 1995a).
We scale the model single star luminosity function to the nearby luminosity
function at $M_{\rm V}\approx10, M_{\rm bol}\approx9$, and plot
$k\,\Psi_{\rm mod,sing}$ (long dashed curve),
$k\,\Psi_{\rm mod,sys}(t=0)$ (dotted curve, without pre-main sequence
brightening) and $k\,\Psi_{\rm
mod,sys}(t=1\,{\rm Gyr})$ (solid curve).
We stress that the solid curves in both panels are luminosity functions for a
realistic model of the Galactic field population of systems consisting of
48~per~cent binaries which have a
period distribution consistent with the empirical G-, K-, and M-dwarf period
distributions, the mass ratio distributions for G-dwarf
systems as observed (Duquennoy \& Mayor 1991), and the overall
mass-ratio distribution given in Kroupa (1995c).
The underlying KTG(1.3) mass function is plotted in Fig.~2.

\vskip 5mm

\noindent {\bf Figure 2.} Open triangles are the stellar mass function we
derive from the luminosity function of Tinney (1993, shown in fig.~4 of Kroupa
1995a) using the log(mass)--bolometric magnitude
relation from Smith (1983) and neglecting binary stars, and the solid dots are
the mass function derived by Scalo (1986). The thick solid line is the
KTG(1.3) mass function (equation~4). It is
the mass function of the model Galactic field star population shown in Fig.~1.

\vskip 5mm

\noindent {\bf Figure 3.} The ideal model single star luminosity function
corresponding to the 95~per
cent confidence range on the KTG($\alpha_1$) mass function index $\alpha_1$
($0.66\le\alpha_1\le1.44$) is
compared with the observational nearby luminosity function (table~2 in Kroupa
1995a) shown as
the histogram and smoothed at the faint end as detailed in section~4 of Kroupa
(1995a).

\vfill
\bye